\documentclass[final]{ias2}
\usepackage{pslatex,multirow,graphicx,array,epsfig}
\def\ra{\rangle}
\def\lr{\leftrightarrow}
\def\mf{m$_{\rm F}$}

\begin{document}
\title{Quantum logic gates using Coherent Population Trapping states}
\author[]{Ashok Vudayagiri}
\email{avsp@uohyd.ernet.in}
\address[{}]{School of Physics, University of Hyderabad\\
Hyderabad 500046}

\begin{abstract}
A scheme is proposed for achieving a Controlled Phase gate using interaction between atomic spin dipoles.
Further, the spin states are prepared in a Coherent Population Trap states, which are robust against
perturbations, laser fluctuations etc. And we show that single qubit and two qubit operations can easily be
obtained in this scheme. The scheme is also robust against decoherences due to spontaneous emissions as the the
CPT states used are dressed states formed out of Zeeman sublevels of ground states of the bare atom. However,
certain practical issues are of concern in actually obtaining the scheme, which are also discussed at the end of
this paper.

\end{abstract}

\maketitle
Conventional computers handle information in the form of bits - which take up
values 0 or 1. `Quantum Computers' on the other hand, use Quantum bits (qubits),
which can be prepared in states 0, 1 or any superposition of the two. Algorithms of quantum computation exploit
this unique feature of quantum mechanical system so as to solve certain class of computational problems with
lesser number of steps. \cite{chuang}. Hence a race to produce a reliable, robust and scalable Quantum Mechanical
system which can be used as gates for Quantum Logic. There have been several attempts in the past to prepare such
a system, using NMR of large molecules, quantum dot structures, ions in linear traps or neutral atoms in Optical lattices
\cite{literature, brennen}, each system withs its own benefits and drawbacks. One of the major requirement
for design of any QC system is that they should be robust and reliable while interactions between any two of them
should be on-demand. One such system is proposed here which involve neutral atoms prepared in  Coherent
Population Trap (CPT) states. It is shown in this paper that such systems can be easily prepared and manipulated
and it is possible to build one-qubit and two-qubit gates using them. Since CPT states are `dark states' of the
atom-light interaction, the atoms prepared in such states will not interact with the light any more
\cite{arimondo, reynaud}. Nor will they evolve in time, since they are already eigenstates of the full
Hamiltonian that consists of atomic as well as interaction terms. 

In this communication, a configuration involving Zeeman sublevels of $^{87}$Rb atom is considered, which exhibits
two different CPT states which can be mapped to two qubits 0 and 1. It is shown that robust states can be
prepared and single qubit and two - qubit operations can be performed using magnetic dipole interactions. 

\section{the configuration}

We consider the transition between $|5S_{1/2}, F=1\rangle$ and $|5P_{1/2}, F=1\ra$ of $^{87}$Rb. \footnote{$|3S_{1/2}, F=1\rangle$ and
$|3P_{1/2}, F=1\ra$ of Na is an equally valid setup with equivalent configuration. We use the dipole-dipole
interactions between one Na and one $^{87}$Rb atom also in a later part of this paper.}, coupled by a two
lasers, which are of same frequency but polarized orthogonal to each other - one in plane containing quantization
axis `z' and other in the 'xy' plane. Following the selection rules \cite{corney} they both couple transitions between
different Zeeman sublevels.

As shown in figure 1b. the beam $E_z= {\cal E}_z\exp[i(\omega t-k.\{x,y\})]$ couples $\Delta m_{\rm F}=0$ transitions
between levels labeled $|g_+\ra \leftrightarrow |e_+\ra$ and $|g_-\ra \leftrightarrow |e_-\ra$. The other beam,
with is plane of polarization in xy plane can be considered as a combination of $\sigma_+$ and $\sigma_-$ beams
coupling $\Delta m_{\rm F}=\pm 1$ transitions $|g_\pm \ra \leftrightarrow |e_0\ra $ and $|g_0\ra
\leftrightarrow |e_\pm \ra$. $|g_0\ra \lr |e_0\ra$ is not coupled by the $E_z$ laser due to
the vanishing Clebsch-Gordon coefficients. Both $E_z$ and $E_p$ beams can be  derived from a same laser source by
use of a halfwave plate and a polarizing beam splitter as shown in figure 1a. The ratio of values of $E_{p,z}$
can be controlled by rotating the halfwave plate HWP.

\begin{figure}
\begin{center}
\psfig{file=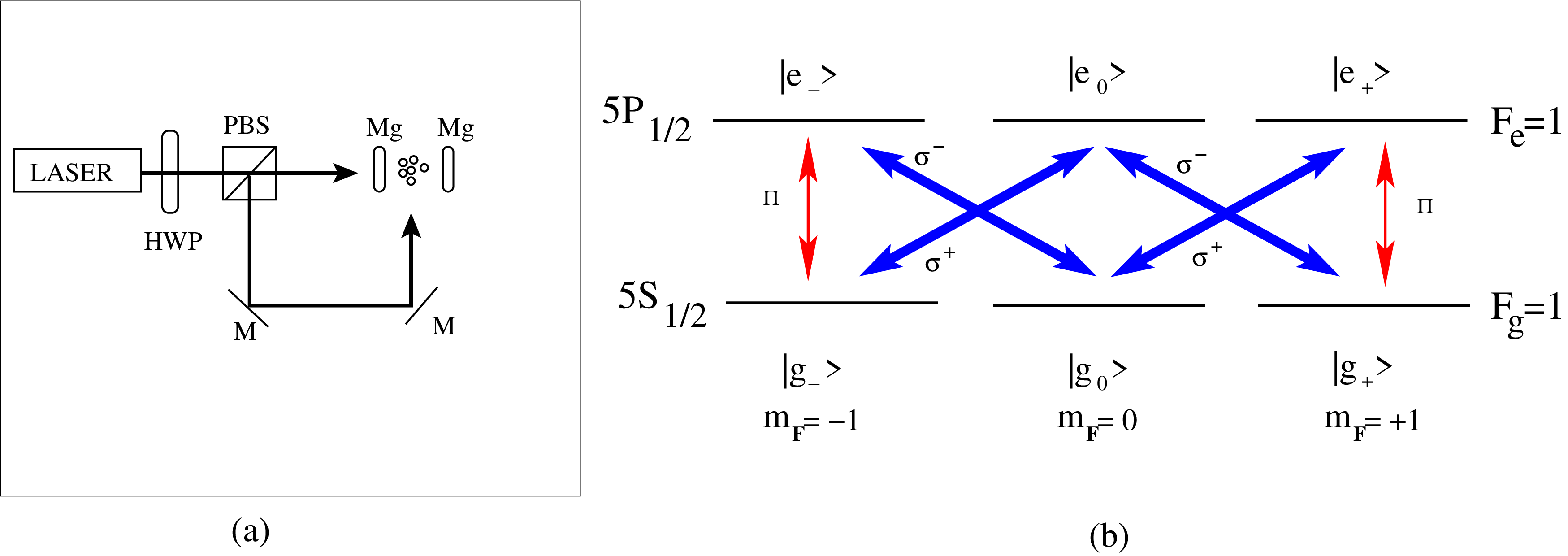,width=8cm}
\end{center}
\caption{(a) Schematic of laser arrangement. HWP is halfwave plate, PBS is polarizing beam splitter and M are
mirrors and Mg are magnets to provide the weak field. (b) Energy level configuration of system used in the setup. Details of the
notations are in detail in the text.}
\end{figure}

When only the $E_p$ beam is present, the configuration is the well known $\Lambda$ system made up of $|g_-\ra \lr
|e_0\ra \lr |g_+\ra$. The steady state solution of this situation is the Coherent Population Trapping (CPT) state
$|\psi_-\ra=(1/\sqrt2)\left[|g_-\ra-|g_+\ra\right]$ \cite{arimondo}. It is interesting to note that $|\psi_-\ra$
is the CPT state even when there exists another CPT configuration - the  V form of $|e_-\ra \lr |g_0\ra \lr
|e_+\ra$, and competes with the $\Lambda$. Our numerical results confirm this fact and it will be shown in a
forthcoming communication. However, in the light of the argument presented in reference \cite{reynaud}, one can
undertand this as a result of atoms trickling from one dressed state to other, eventually reaching the the state
$|\psi_-\ra$. On the other hand, when only $E_z$ beam is present, then all the atoms in $|g_\pm\ra$ will be optically pumped
out and eventually reach $|g_0\ra$. This is a trap state for the $E_z$ beam. The two trap states $|\psi_0\ra=|g_0\ra$ and  $|\psi_-\ra$ can now be mapped to the qubit states
$|\psi_0\ra=|0\ra$ and $|\psi_-\ra=|1\ra$. 

More interestingly, if both  $E_p$ and $E_z$ beams are present together, the steady state solution then is not a
statistical mixture of the two trap states $|\psi_0\ra$ and $|\psi_-\ra$, but a three component CPT states 
\cite{ashokspt}   

\begin{equation}
|\psi \rangle = \frac{\left(\Omega_p/\Omega_z\right)|g_o \rangle - |g_- \rangle + |g_+ \rangle}{\sqrt{ 2 +
|(\Omega_p/\Omega_z)|^2}} \label{threetrap},
\end{equation}

which can be rewritten as 

\begin{eqnarray}
|\psi\ra&=&\sin (\theta/2) |\psi_0\ra +\exp(i\phi)\cos (\theta/2) |\psi_-\ra. \\ \nonumber
{\rm Or}&&\\
|\psi\ra&=&\sin (\theta/2) |0\ra +\exp(i\phi)\cos (\theta/2) |1\ra, 
\label{state2}
\end{eqnarray}

where 
\begin{equation}
\sin (\theta/2)=\frac{\Omega_p}{\sqrt{2|\Omega_z|^2+|\Omega_p|^2}} ~~~~{\rm and} ~~~~\cos (\theta/2)
=\frac{(\sqrt{2}\Omega_z)}{\sqrt{2|\Omega_z|^2+|\Omega_p|^2}}
\label{rabi}
\end{equation}

and $(\theta/2)=\tan^{-1}\left(\Omega_p/\sqrt2\Omega_z\right)$. Any desired value of $\theta$ can be obtained by
varying the ratio of $\left(\Omega_p/\sqrt2\Omega_z\right)$, where $\Omega_{p,z}=d.E_{p,z}/2\hbar$. The phase
factor $\phi$ in (\ref{state2}) can also be obtained by controlling the phase between the two beams
$E_{p,z}={\cal E}_{p,z} \exp[i(\omega t-k.x-\phi_{p,z})]$.  If the setup is as in figure 1a, then rotating the
HWP will distribute the intensity between $E_p$ and $E_z$ and will positioning it appropriately will produce any
desired $\theta$.  Keeping a variable retarder at one of the output ports of PBS will also control the phase
$\phi$.

The operation then can be mathematically expressed by
\begin{equation}
H(\theta)=\left(\begin{array}{cc}\sin(\theta/2) & e^{i\phi}\cos(\theta/2) \cr
-e^{-i\phi}\cos(\theta/2) &
\sin(\theta/2)\end{array}\right) \label{hadamard},
\end{equation}

which, acting on the basis vectors 
\begin{equation}
|0\ra \equiv \left(\begin{array}{c}1 \\ 0 \end{array} \right)~~~{\rm and}~~~|1\ra \equiv\left(\begin{array}{c}0 \\ 1
\end{array} \right),
\label{basis}
\end{equation}

leads to dressed state vectors
\begin{equation}
|\Phi_-\ra \equiv \left(\begin{array}{c} \sin(\theta/2)\\ -e^{-i\phi}\cos(\theta/2) \end{array}
\right)~~~{\rm and}~~~|\Phi_+\ra \equiv \left(\begin{array}{c} e^{i\phi}\cos(\theta/2)\\ \sin(\theta/2)
\end{array} \right).
\label{dressed}
\end{equation}

Operator $H(\theta)$,  reduces to a Hadamard when $\theta/2$ is set to 45$^{0}$
and $\phi=1$, which is equivalent of setting the halfwaveplate of fig. 1a to at
45$^{0}$. The state $\Phi_+$is is a CPT state, as given in equation
(\ref{state2})

This scheme for state preparation has certain distinct advantages.  (i) The
qubit states (\ref{basis}) as well as state $\Phi_+$ are CPT states. CPT states are end points of atom-laser
interaction and the atoms eventually reach CPT states via non-CPT states as shown by Cohen-Tannoudji and Reynaud
\cite{reynaud}. This means that the state preparation is reliable and the desired state is always prepared. (ii)
Once the states are prepared, the atoms in these state no longer interact with the laser that prepares them. This
eliminates need for precise time-control of the lasers. The state preparation is therefore robust and certain.
(iii) The state preparation involves only cw beams and does not require any complex
pulse shaping schemes. (v) Since it does not involve single photon processes, lasers with nominally high
intensity can be used. This would allow very precise control of phase $\phi$ while allowing fluctuations in the
intensity. (vi) Any desired superposition corresponding to any desired Bloch vector can be prepared by simply
varying the intensity ratio between two laser beams.  Due to all of above, the configuration allows a robust and
reliable preparation of two qubit states and its superposition and also the method of state preparation is very easy. In
the following sections, methods of performing single qubit and two-qubit operations are discussed.  

\section{Operations of logic gates}
\subsection{Single Qubit Operations}
Setting $\theta/2=0$ in (\ref{hadamard}) will result in a rotation, which is the NOT operation
 
\begin{equation}
H(\theta=0)=\left(\begin{array}{cc}0 & e^{i\phi} \cr
-e^{-i\phi}  & 0\end{array}\right) \label{rotation},
\end{equation}

which converts $|\psi\ra=\sin (\theta/2) |0\ra +\exp(i\phi)\cos (\theta/2) |1\ra$ to
$|\psi\ra=\cos (\theta/2) |0\ra +\exp(i\phi)\sin (\theta/2) |1\ra$, for any value of existing $\theta/2$.

This is an intriguing situation since setting $\theta/2=0$ in equation (\ref{hadamard}) is equivalent to
setting $\Omega_p=0$ in (\ref{rabi}), which is equivalent to switching off $E_p$ beam and thus always creating
the atoms in state $|1\ra$, no matter what is the original state. This discrepancy can be understood in the
manner that the NOT operation always operates on the full dressed state $|\Phi_+\ra$ and hence valid.

\subsection{Two-qubit operation; C-Phase gate}
Two-qubit opeations can be obtained in a manner similar to the earlier works that exploited the dipole-dipole interaction \cite{ryabtsev, brennen}, except using magnetic
dipole-dipole interaction between spin states instead of electric dipoles. 

In an external magnetic field $\vec B$, the spin vectors align at an angle that depends on their \mf value and
also makes a Larmor precision about $\vec B$, with a frequency $\omega_L= \gamma_L |B|$. $\gamma_L$ is the
gyromagnetic ratio of the atom and $|B|$ is the value of the magnetic field. The atom can now be flipped from one \mf
state to the other by applying an oscillatory magnetic field perpendicular to $\vec B$, and at a frequency equal
to difference between the two corresponding Larmor frequenices. The dipole-dipole interaction between the spins
now manifest as a shift in the Larmor frequencies and hence the resonance frequency for the oscillatory magnetic
field also shifts as shown in figure 2. \cite{cohentannoudji,jackson}

As in case of electric dipoles, the spin dipole interaction is also inversely dependent on the cube of the
distance between them, given by 

\begin{equation}
V_{dd}=\frac{\mu_0}{4\pi}\frac{\gamma_L^2}{r^3}\left[S_1.S_2-3(S_1.n)(S_2.n)\right].
\label{dipoledipole}
\end{equation}
Which gets reduced to $V_{dd}=\frac{\mu_0}{4\pi}\frac{\gamma_L^2}{r^3}\left(3\cos^2\theta_s-1\right),$ for two degenerate $m_{\rm F}$
levels\cite{cohentannoudji}.

Here $r$ is the normal distance between the two atoms, $\theta_s$ the angle between the spin directions and $\vec
r$, $\mu_0$ the permittivity of free space and the ratio $\mu_0/4\pi$ is a scaling factor for MKS units. The
energy levels of the states atom pairs can be shown as in figure 2a. This interaction $V$ causes a mixing of
the pair states $|01\ra$ and $|10\ra$ as well as cause a shift in the energies as shown in figure 2 b. The
energy for the transition $|00\ra \lr |10\ra + |01\ra$ is shifted by $\Omega_m=2 \left(\mu_0/(4\pi)\right).\left(\gamma_L^2/{r^3}\right)$ 

\begin{figure}
\begin{center}
\psfig{file=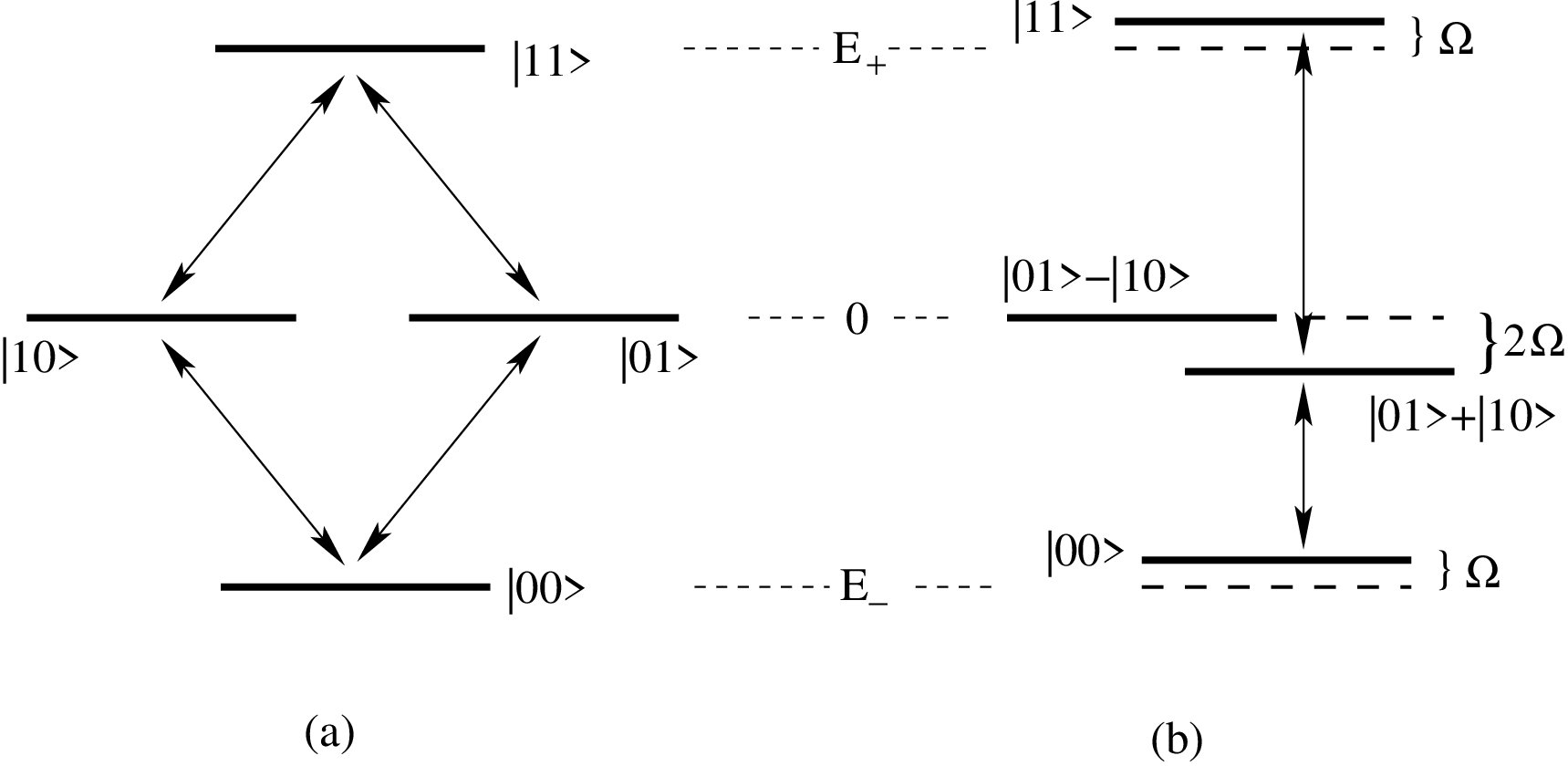,width=7cm}
\end{center}
\caption{(a) Energy diagram of two-atom system (b) shows effect of the dipole-dipole interaction. The state $|10\ra -|01\ra$ is not
coupled by radio frequency transition either to $|00\ra$ or to $|11\ra$. The energy difference in case of
non-interaction situation is equal to $\hbar \omega_L$ where $\omega_L$ is the Larmor frequency. See text for
amount of shift in case of (b). }
\end{figure}

If a pulse of oscillatory magnetic field at frequency $\omega_L+2\Omega$ the atoms will absorb its energy only if
they are in the state $|10\ra+|01\ra$, and travel to state $|11\ra$. If such a pulse has a McCall-Hahn pulse area
of 2$\pi$, the the atom will return to state $|10\ra+|01\ra$, but now with a phase factor of $\pi$. Atoms in any
other state will not be affected. If the atom pair is in state $|10\ra-|01\ra$ instead, the phase factor already exists
for $|01\ra$ state.  Therefore, if the atoms are brought together, the rf pulse applied and then
taken apart, only the atoms in state $|01\ra$ will return to $-|01\ra$. 

Another option, following Ryabstev and co-workers \cite{ryabtsev} is to bring the atoms together and hold them
close for a specific period. Since the dipole-dipole interaction causes a mixing of the states $|01\ra$ and
$|10\ra$ for form a time dependent superposition $$|\psi_{dd}(t)\ra = \cos (V_{dd}t/\hbar)|10\ra -
i~\sin(V_{dd}t/\hbar)|01\ra.$$, the atom pair oscillates between $|01\ra$ and $|10\ra$ with a half-period
$T=1/2 (\hbar \pi)/V_{dd}$. The scheme then involves holding the atoms together for $T$ and taking apart, which
gives a control-swap gate or for 2$T$, which will result in a controlled swap gate. 

The serious drawback of this scheme is that one can not distinguish a-priori between control atom and the logic
atom since they are both identical. An option then is to use two different atoms, with same level configuraiton.

\subsection{heterogeneous atoms} Sodium, with $3S_{1/2},F=1$ and $3P_{1/2}, F=1$ triplets, shows an identical
behaviour of state preperation and qubit operations, but the corresponding Larmor frequency is different. The
dipole-dipole interaction between spins of Sodium atom and $^{87}$Rb will cause only a level shift instead of a
mixing states $|01\ra$ and $|10\ra$ as showin figure 3 \cite{cohentannoudji}. The amount of shift is 
$\Omega=(\mu_0/4\pi)(\gamma_1\gamma_2/r^3)(3\cos^2\theta_s-1)$, where $\gamma_1$ and $\gamma_2$ are gyromagnetic
ratios of the Sodium and $^{87}$Rb atoms respectively. 

\begin{figure}
\begin{center}
\psfig{file=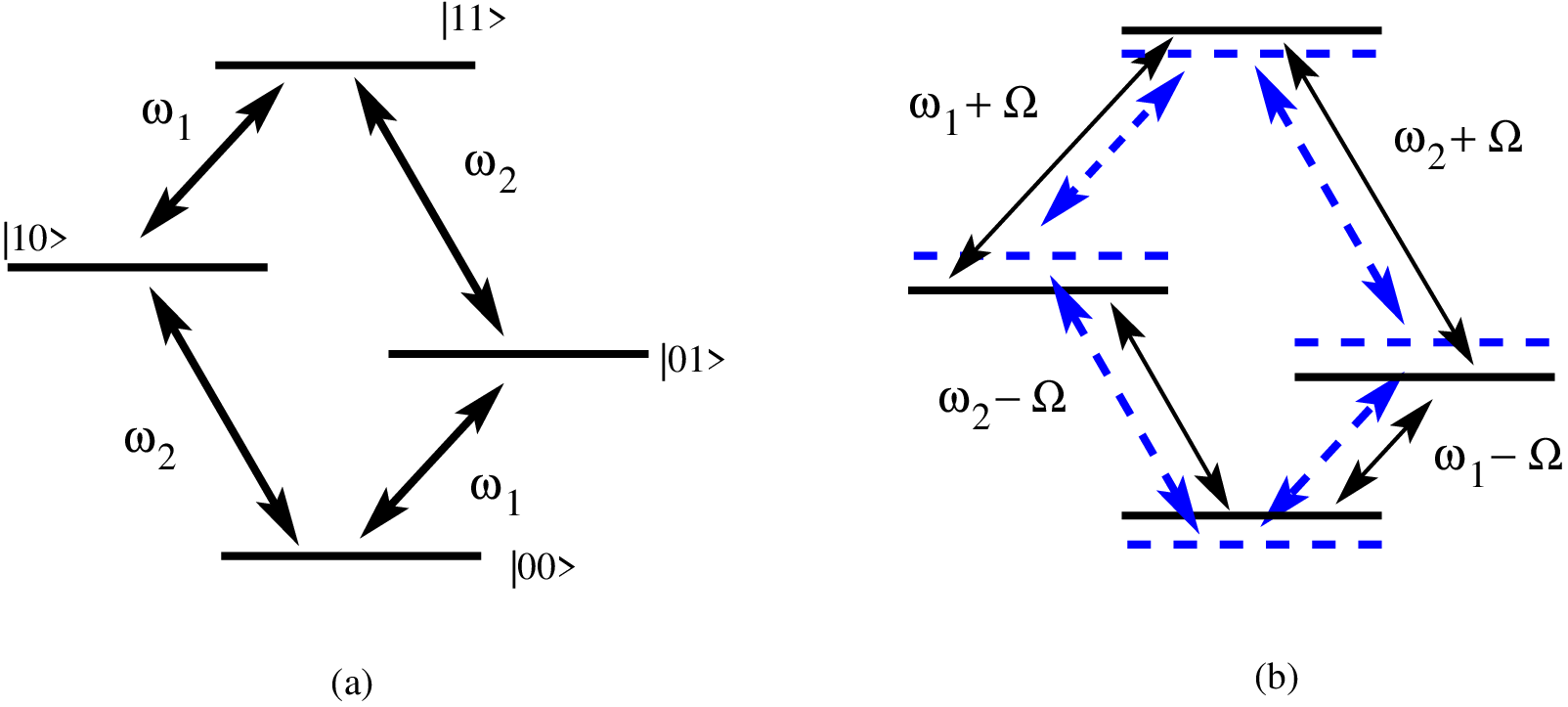,width=7cm}

\end{center}
\caption{Energy diagram of two-atom system for two different atoms. (a) without the spin
dipole-dipole interaction and (b) Energy shifts due to dipole interaction. Note that there is
no mixing. dashed lines  in (b) are position of unshifted energy levels. }
\end{figure}

Now a controlled NOT gate can be obtained by using a rf pulse with a frequency $\omega_2+\Omega$ with a
McCall-Hahn area of 2$\pi$, or a pair of pulses with frequencies $\omega_2+\Omega$ and $\omega_1+\Omega$ times in
a STIRAP like fashion to obtain a controlled swap gate.

\section{conclusion}  
It is shown that a system that exhibits two trapping states can be obtained in  $^{87}$Rb and Sodium atom
interacting with two lasers that couple its F=1 $\lr$ F=1 transition. They can be mapped to the qubit states
$|1\ra$ and $|0\ra$ and can be used for quantum computation. Any required superposition state
$|\psi\ra=\sin (\theta/2) |0\ra +\exp(i\phi)\cos (\theta/2) |1\ra$ can be prepared. Since this involves CPT
states and also ground levels, it is very robust against decoherences. Single qubit and two-qubit operations are
described with these states. However, the dipole-dipole interaction between spin states is weaker than that
between electric dipole states and hence the shift is small. But the typical value of Larmor frquency for most
alkali atoms are about a 100 kHz and measuring a small shift in the Radio frequency domain is technologically
feasible. The major technical difficulty with this scheme may be with having to move the atoms nearer and apart
as and when required for the two qubit operation. 

\section{acknowledgment}
Several useful discussions led to fine tuning of this work. In particular, I thank Hema Ramachandran of Raman
Institute and R. Srikanth of Poorna Prajna Institute, Bangalore, Prasanta Panigrahi of IISER, Kolkota and Surya
P. Tewari of University of Hyderabad, for helpful discussions.

\end{document}